\documentstyle[epsfig, twocolumn]{esapub}

\newcommand\be{\begin{equation}}
\newcommand\ee{\end{equation}}
\newcommand\ga{{\leavevmode\kern0.3em\raise.3ex\hbox{$>$}
\kern-0.8em\lower.7ex \hbox{$\sim$}\kern0.3em}}
\newcommand\la{{\leavevmode\kern0.3em\raise.3ex\hbox{$<$}
\kern-0.8em\lower.7ex \hbox{$\sim$}\kern0.3em}}

\begin{document}

%% Do not remove the following six lines:
\setlength{\parindent}{0pt}
\setlength{\parskip}{ 10pt plus 1pt minus 1pt}
\setlength{\hoffset}{-1.5truecm}
\setlength{\textwidth}{ 17.1truecm }
\setlength{\columnsep}{1truecm }
\setlength{\columnseprule}{0pt}
\setlength{\headheight}{12pt}
\setlength{\headsep}{20pt}
\pagestyle{esapubheadings}

%% Title - should be in capitals:
\title{\bf RING DIAGRAM ANALYSIS OF VELOCITY FIELDS WITHIN THE SOLAR
CONVECTION ZONE}

%% If the author list spans more than one line then the {\bf (bold
%% font)} command must be inserted for each line
\author{{\bf Sarbani Basu$^1$, H. M. Antia$^2$, S. C. Tripathy$^3$} \vspace{2mm} \\
$^1$Institute for Advanced Study, Olden Lane, Princeton N. J. 08540, U. S. A. \\
$^2$Tata Institute of Fundamental Research,
Homi Bhabha Road, Mumbai 400005, India \\
$^3$Udaipur Solar Observatory, Physical Research Laboratory,
PO Box No. 198, Udaipur 313 001, India}

\maketitle

\begin{abstract}

Ring diagram analysis of solar oscillation power spectra obtained from
MDI data is performed to study the velocity fields within the
solar convection zone. The three dimensional power spectra are fitted
to a model with a  Lorentzian profile in frequency and includes the advection
of the wave front by horizontal flows to obtain the two horizontal
components of flows as a function of the horizontal wave number and
radial order of the oscillation modes. This information is then inverted
using the OLA and RLS techniques to infer the variation in flow velocity
with depth. The resulting velocity fields yield the mean rotation velocity
at different latitudes which agrees reasonably with helioseismic estimates.
The zonal flow inferred in the outermost layers also appears to be
in agreement with other measurements.
A meridional flow from equator polewards is found to have
an amplitude of about 25 m/s near the surface and the amplitude appears to
increase with depth.
\vspace {5pt} \\

%% Do not remove the previous commands. Your abstract should 
%% end with \vspace {5pt} \\  

%% Please insert your keywords here.
  Key~words: Sun: oscillations; Sun: rotation; Sun: interior.

\end{abstract}

\section{INTRODUCTION}

The rotation rate in the solar interior has been inferred using
the frequency splittings for p-modes (Thompson et al.~1996;
Schou et al.~1998). However, these splitting coefficients of the
global p-modes are sensitive only to the North-South axisymmetric
component of rotation rate. To study the nonaxisymmetric component of
rotation rate and the meridional component of flow, other techniques
based on `local' modes are required.
Since these velocity components are comparatively small in magnitude
they have not been measured very reliably even at the solar surface.
Apart from these nearly steady flows there could also be cellular flows with
very large length scales and life-times, i.e., the giant cells,
which have been believed to exist, although there has been no firm
evidence for such cells (Snodgrass \& Howard 1984; Durney et al. 1985;
Howard 1996).
Study of such large scale flows is important
for understanding the theories of solar dynamo and turbulent compressible
convection (Choudhuri et al.~1995; Brummell et al.~1998).

The high degree modes ($l\ga150$)
which are trapped in the solar envelope have lifetimes which are much
smaller than the sound travel time around the Sun and hence the
characteristics of these modes are mainly determined by localized average
conditions rather than average over entire spherical shell. These modes
can  be employed to study large scale flows inside the Sun, using
the time-distance helioseismology (Duvall et al.~1993, Giles et al.~1997),
Ring diagrams (Hill 1988; Patron et al.~1997) and other techniques.

Ring diagram analysis is based on the study of three-dimensional  power 
spectra of solar
p-modes on a part of the solar surface. If we consider a section of this
3d spectrum at fixed temporal frequency, the power is concentrated
along a series of rings that  correspond to different values of the
radial harmonic number $n$. The frequencies of these modes are also
affected by horizontal flow fields suitably averaged over the region under
consideration. Hence, an accurate measurement of these frequencies
will contain the signature of large scale flows and can be used
to study these flows. Since the high degree modes used in these studies
are trapped in the outermost layers of the Sun, such analysis gives
information about the conditions in the outer 2-3\% of the solar radius.

\section{THE TECHNIQUE}

In this work we have used data from full-disk Dopplergrams obtained
by the SOI/MDI instrument on board SoHO. Selected regions of
Dopplergrams mapped with Postel's projection are tracked at a rate
corresponding to the photospheric rotation rate at the center of
each region to  filter out the
photospheric rotation velocity from the flow fields. This allows
us to study the smaller components of the flow which are not very
well determined from other studies.
For each tracked region, the images are
detrended by subtracting the running mean over 21 neighboring
images to filter the series temporally. The
detrended images are apodized and
Fourier transformed in the two spatial coordinates and in time to
obtain the 3d power spectra. We have chosen the spatial extent of
the region to be about $15^\circ\times15^\circ$ with $128\times128$ pixels
in heliographic longitude and latitude giving a resolution of
$0.03367$ Mm$^{-1}$ or 23.437 $R_\odot^{-1}$. Each region is tracked for
4096 minutes giving a frequency resolution of 4.07 $\mu$Hz.
To minimize effects of foreshortening all the regions were centered
on the central meridian.
The spectra have been obtained using the usual tasks in MDI
data processing pipeline.
Fig.~1 shows some of the sections at constant $\nu$ of these spectra.

\begin{figure}[t]
  \begin{center}
    \leavevmode
  \centerline{\epsfig{file=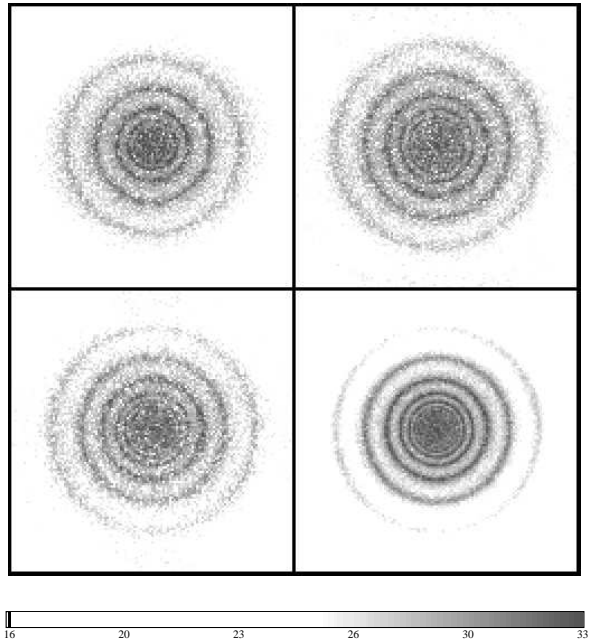,width=8cm}}
  \end{center}
  \caption{\em 
Sample (logarithmic) power spectra
as a function of $k_x$ and $k_y$
at a fixed frequency. The top panels are for the region centered at
equator and Carrington longitude of $60^\circ$, at frequencies of
around 3 mHz (left) and 4 mHz (right). The bottom left panel is
the spectrum around 4 mHz for a region centered at $40^\circ$N latitude
and longitude of $60^\circ$. The bottom right panel is the summed spectrum
around 4 mHz for regions centered at the equator. The gray scale is
marked with logarithm of the power.
}
\end{figure}

To extract the flow velocities and other mode parameters from the
3d power spectra we fit a model of the form
\begin{eqnarray}
\lefteqn{P(k_x,k_y,\nu)={e^{B_2}\over k^2}+{e^{B_3}\over k^3}+
\qquad\qquad\qquad}\nonumber \\
& \qquad\qquad{\exp(A_0+(k-k_0)A_1+A_2({k_x\over k})^2+
A_3{k_xk_y\over k^2})\over
(\nu-ck^p-U_xk_x-U_yk_y)^2+w^2}
\end{eqnarray}
where $k^2=k_x^2+k_y^2$, and the 11 parameters $A_0, A_1, A_2, A_3, c, p,
U_x, U_y, w, B_2$ and $B_3$ are determined by fitting the spectra
using a maximum likelihood approach
(Anderson et al.~1990). Here $k_0$ is the central value of $k$ in the
fitting interval.
The $A_2$ and $A_3$ terms account for the variation of power along the
ring.
The minimization has been performed using a
quasi-Newton method based on the BFGS formula for updating the Hessian
matrix (Antia 1991).
The form given by Eq.~(1) is slightly different from what
is used by Patron et al.~(1996), because we have assumed some variation
in amplitude along the ring, coming from the $A_2$ and $A_3$ terms.
These were introduced because the power does appear to vary along
the ring and the fits in the absence of these terms were not satisfactory.
%In fact, we find that even these terms are not sufficient to account
%for power variation in spectra at high latitudes possibly due to
%foreshortening.

We fit each ring separately by using the portion of power spectrum
extending halfway to the adjoining rings. For each fit we use
a region extending about $\pm100\mu$Hz from the chosen central frequency.
We choose the central frequency for fit in the range of 2--5 mHz
as the power outside this range  is not significant. The
rings corresponding to $0\le n\le6$ have been fitted.
This gives us typically 800 `modes', all of them may not be independent
as there is a considerable overlap between adjacent fitting intervals.

The fitted $U_x$ and $U_y$ for each mode represents an average
over the entire region in horizontal extent and over the vertical
region where the mode is trapped.
We can invert the fitted $U_x$ (or $U_y$) to infer the variation in
horizontal flow velocity $u_x$ (or $u_y$) with depth.
We use the Regularized Least
Squares (RLS) as well as  the Optimally Localized Averages (OLA)
techniques for inversion. For the purpose of inversion the fitted
values of $U_x$ and $U_y$ are interpolated to the nearest integral
value of $k$ (in units of $R_\odot^{-1}$) and then the kernels
computed from a full solar model with corresponding value of degree
$\ell$ are used for inversion.

We have selected the regions centered at Carrington longitudes of
$90^\circ, 60^\circ, 30^\circ$ for rotation 1909 and at
$360^\circ, 330^\circ, 300^\circ$ for rotation 1910 corresponding
to period from about May 24 to June 7, 1996. For each longitude
we select regions centered at latitudes of
$0^\circ$, $\pm10^\circ$,
$\pm20^\circ$, $\pm30^\circ$, $\pm40^\circ$, $\pm50^\circ$ and
$\pm60^\circ$.
There is some overlap between different regions.
Apart from these individual spectra we have also analyzed spectra obtained
by summing all 6 spectra for a given latitude to study the properties
averaged over different longitudes. Because of averaging, these spectra
have better statistics and the error estimates are also lower.
These averaged spectra can be expected to give the average velocity
over the range of longitudes considered. Most of the inferences in this
work have been obtained using these averaged spectra.
We fit each spectrum to obtain the mode parameters
including $U_x$ and $U_y$.

\section{RESULTS}

Following the procedure outlined above we fit the form given by Eq.~(1)
to a suitable region of a  3d spectrum.
Although other quantities may also be of some interest, in this work we
restrict our attention to the two horizontal components of velocity (Fig.~2)
obtained by fitting the spectra.
The fitted velocities for each `mode' are then inverted to obtain the
variation of horizontal velocity with depth. Only the region $r>0.97R_\odot$
is sampled by the modes used in this study and hence the inversions are
restricted to this region. Some of the results obtained using the
RLS and OLA techniques are shown in Fig.~3.

\begin{figure}[!ht]
  \begin{center}
    \leavevmode
  \centerline{\epsfig{file=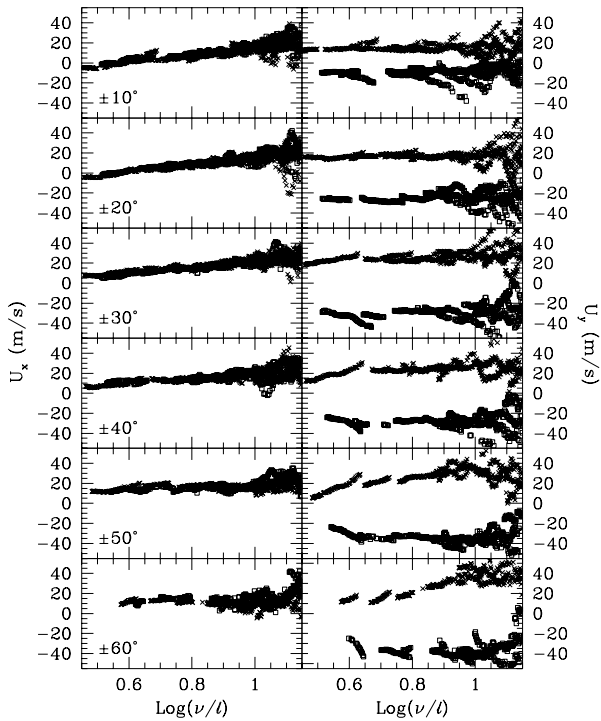,width=8cm}}
  \end{center}
\vskip -0.4 cm
  \caption{\em  The fitted velocity for summed power spectra at
different latitudes. In each panel the crosses mark the fitted
velocity for northern hemisphere while the open squares mark that for
the southern hemisphere. In the right panels the crosses marking
northern latitudes generally fall in the upper half of the figure, while the
squares marking southern latitudes fall in the lower half.
The latitudes are marked in the left panel.
Error bars are not shown for clarity.
}
\end{figure}

\begin{figure}[!ht]
  \begin{center}
    \leavevmode
  \centerline{\epsfig{file=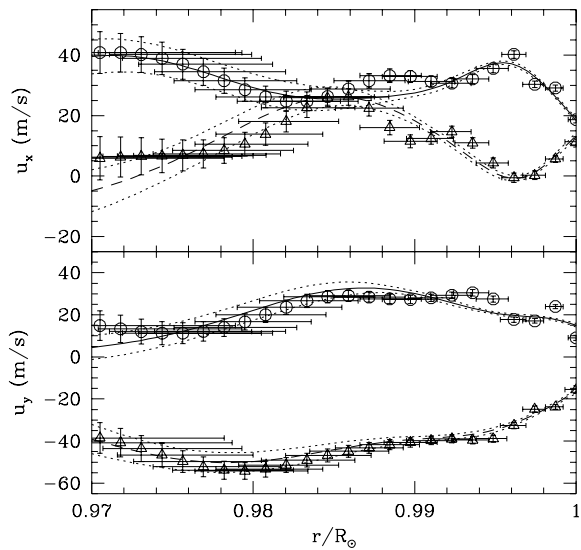,width=8cm}}
  \end{center}
\vskip -0.4 cm
  \caption{\em  A sample set of inversions are shown for latitudes
%? Is it 330 or 300? The file name says 300! -- A
$\pm 40^\circ$ and longitude $300^\circ$. The circles and triangles are
the OLA inversions for $40^\circ$N  and $40^\circ$S respectively. The
continuous and dashed lines are the RLS results for
$40^\circ$N  and $40^\circ$S respectively, with the dotted lines denoting
the $1\sigma$ error limits.
}
\end{figure}

From the inversion results it appears that the longitudinal component
($u_x$) is dominated by the average rotational velocity.
This is a result of the fact that tracking is done at the surface rotation
rate at the centre of the tracked region and hence does not account for the
variation of the rotation rate with depth.
The average variation with latitude, $\theta$ of these components
can be determined from the summed spectra and a reasonable agreement between
$u_x$ and the rotation rate determined from splitting coefficients of the
global p-modes can
be seen from Fig.~4. The rotation velocity at each latitude can be
decomposed into the symmetric part [$(u_N+u_S)/2$] and an
antisymmetric part [$(u_N-u_S)/2$]. The symmetric part can be compared
with the rotation velocity as inferred from the splittings of global modes
(Antia et al.~1998) and the results are shown in Fig.~5.
Since the inversion results using global modes are not
particularly reliable in the surface regions, the velocity profiles obtained
from the ring diagram analysis supplement those results and support the
earlier conclusions. 

\begin{figure}[t]
  \begin{center}
    \leavevmode
  \centerline{\epsfig{file=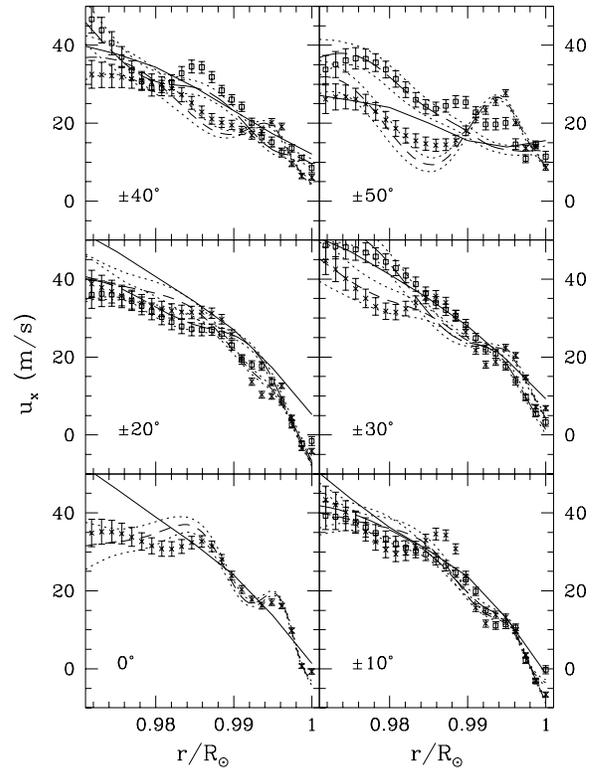,width=8cm}}
  \end{center}
  \caption{\em  The average horizontal velocity at
different latitudes are
compared with the rotation velocity (after subtracting out the
surface rotation rate used in tracking each region)
obtained from inversion of splitting-coefficients (continuous line).
The results obtained using RLS inversions are shown by short-dashed
(northern latitudes)
and long-dashed lines (southern latitudes) with dotted lines marking the
$1\sigma$ error limits. Similarly, the crosses (north) and squares (south)
with error bars represent the results of OLA inversions.
}
\end{figure}

\begin{figure}[!ht]
  \begin{center}
    \leavevmode
  \centerline{\epsfig{file=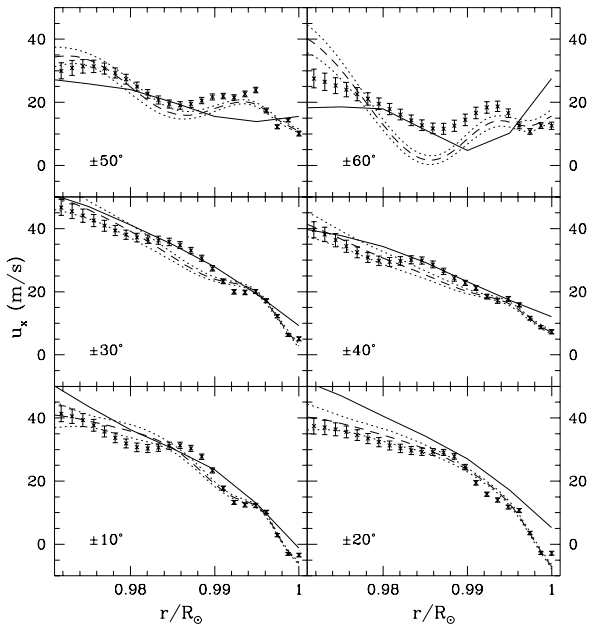,width=8cm}}
  \end{center}
  \caption{\em  
The average horizontal velocity at
different latitudes (dashed line for RLS and crosses for OLA),
i.e., the latitudinally symmetric part [$(u_N+u_S)/2$],
compared with the rotation velocity (after subtracting out the
surface rotation rate used in tracking each region)
obtained from inversion of splitting-coefficients (continuous line).}
\end{figure}

Following Kosovichev \& Schou (1997) it is possible to decompose the
rotation velocity into two components, a smooth part [polynomial in
terms of $\cos(\theta)$, $\cos^3(\theta)$ and $\cos^5(\theta)$]
and the remaining part which has been identified with zonal flows.
The zonal flow so estimated is shown in Fig.~6. Despite poor latitudinal
resolution in our results the inferred pattern  near the surface
is in reasonable agreement with the average zonal flow estimated from the
splitting coefficients for the f-modes from the 360 day MDI data.
But
at deeper depths the pattern changes significantly and the errors are
also larger. Hence, it is not clear if
the zonal flow penetrates below about 7Mm ($0.01R_\odot$) from the surface.

\begin{figure}[!ht]
  \begin{center}
    \leavevmode
  \centerline{\epsfig{file=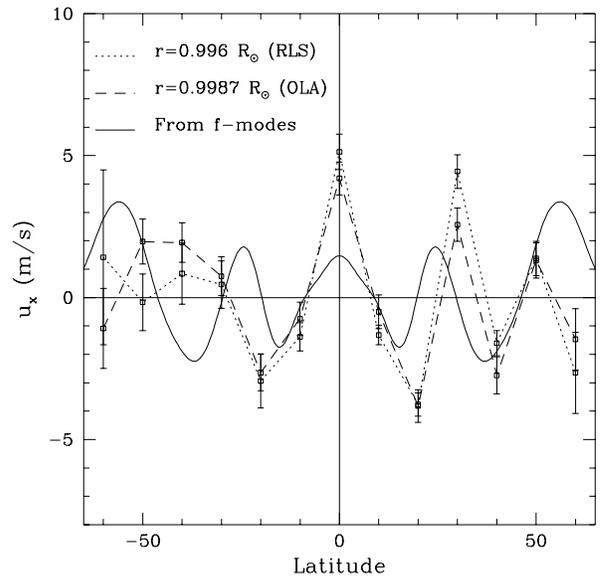,width=8cm}}
  \end{center}
  \caption{\em  The zonal flow, i.e., the rotation velocity
after removing a smooth component, at a layer just below the solar
surface. The dotted line and points are RLS results, while the dashed
line and points are OLA results. The error bars shown in this figure
represent the error in rotation velocity and do not include any
contribution from the smooth component which is subtracted to obtain
these values. Hence the errors are likely to be underestimated.
The continuous line represents the average zonal flow velocity as inferred
from the f-modes using the 360 day MDI splitting coefficients.}
\end{figure}

\begin{figure}[!ht]
  \begin{center}
    \leavevmode
  \centerline{\epsfig{file=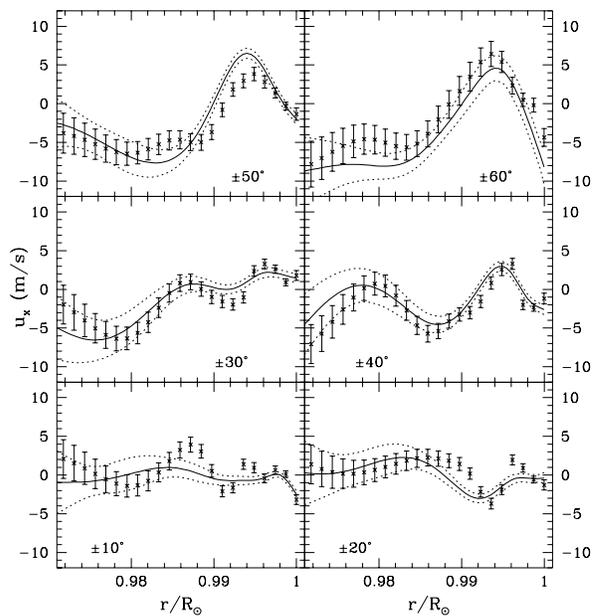,width=8cm}}
  \end{center}
  \caption{\em  The antisymmetric component [$(u_N-u_S)/2$] of the
rotation velocity plotted as a function of depth for
various latitudes. The lines are RLS and crosses are OLA results.}
\end{figure}

The difference between rotation velocity at the
same latitude in the North and South hemispheres is small and thus
the antisymmetric component of rotation rate is not very significant.
Some of the
difference may also be due to some systematic errors in our analysis.
For example, due to differences in the angle of
inclination for the regions at same latitude in the two hemisphere, the
effect of foreshortening will be different in the two hemispheres.
The antisymmetric component of the velocities is shown in
Fig.~7. In particular, it can be seen that at low latitudes where the
results are more reliable, the antisymmetric component is rather small,
more or less within error estimates.

\begin{figure}[!ht]
  \begin{center}
    \leavevmode
  \centerline{\epsfig{file=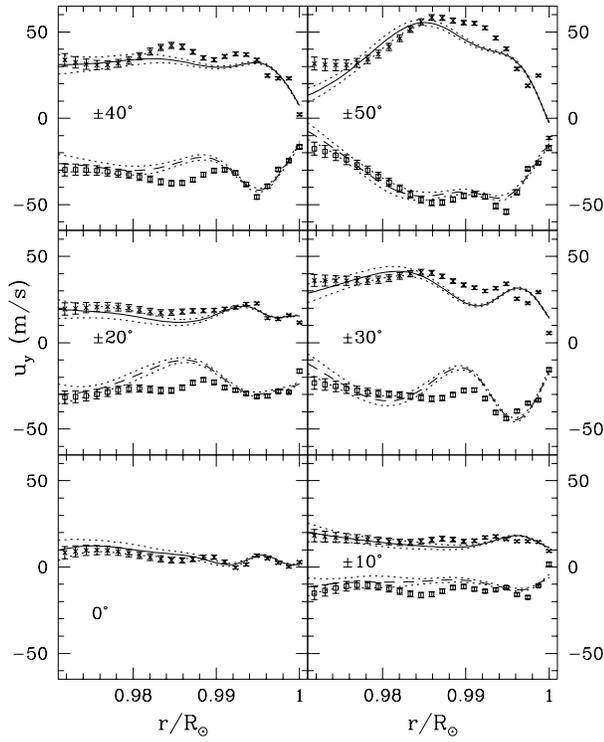,width=8.0cm}}
  \end{center}
\vskip -0.2 cm
  \caption{\em  Meridional velocity at different latitudes
plotted as a function of depth.
The results obtained using RLS inversions are shown by continuous
lines (northern latitudes)
and dashed lines (southern latitudes) with the dotted lines showing the
$1\sigma$ error limits. Similarly, the crosses (north) in the positive
half and squares (south) in the negative half of each panel
represent the results of OLA inversions.
}
\end{figure}

\begin{figure}[!ht]
  \begin{center}
    \leavevmode
  \centerline{\epsfig{file=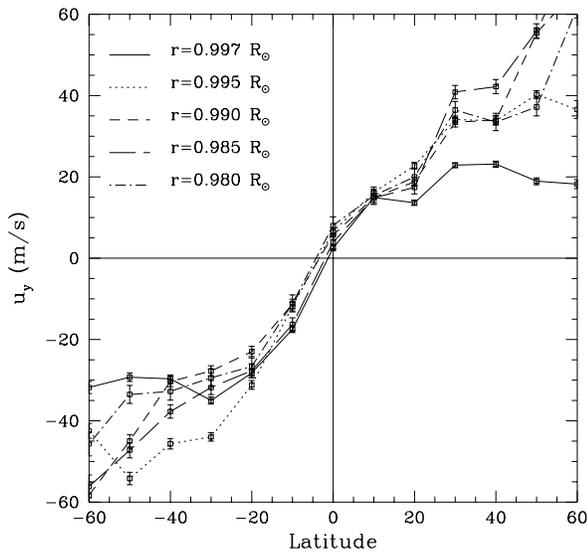,width=8.0cm}}
  \end{center}
\vskip -0.2 cm
  \caption{\em  Meridional velocity at different depths
plotted as a function of latitude. These results have been obtained
using the  OLA technique for inversion.
}
\end{figure}

The latitudinal component of the velocity appears to be dominated by the
meridional flow from equator polewards. The average latitudinal velocity
for each latitude is shown in Fig.~8, while Fig.~9 shows the same
 as a function of latitude at a few selected depths.
There is a significant variation in this velocity with depth at high latitudes.
Since the measurements are not particularly
reliable at high latitudes it is difficult to say much about the
general form of the flow velocity with latitude and moreover it will
also depend on depth. However, if we fit to
a form (cf., Giles et al.~1997)
\be
u_y=a_1\cos(\theta)+a_2\cos(2\theta)
\ee
then we find mean flow $a_2$ of about 25--40 m/s depending on the
depth, while $a_1$ is small, more
or less comparable to error estimates. It is not clear if the
assumed form indeed fits the measured variation as the resulting
$\chi^2$ is fairly large. The expected decrease in the magnitude of
$u_y$ at high latitudes, is not very clear from the results,
though at layers immediately below the surface the meridional
velocity does appear to decrease with latitude at high latitudes.
The amplitude is about 25 m/s at a radial distance of $0.997R_\odot$
which is comparable to the estimated value obtained by Giles et al.~(1997)
from time-distance analysis and by Hathaway et al.~(1996) from direct
Doppler measurement at solar surface. The amplitude appears to increase
rapidly with depth to about 40 m/s, and it is not clear if this form of
latitudinal variation is actually valid at deeper depths where the velocity
does not
show any turnover at high latitudes. There is no evidence for any change
in sign of the meridional velocity up to a depth of $0.03R_\odot$ or
21 Mm that is covered in this study.

\begin{figure}[t]
  \begin{center}
    \leavevmode
  \centerline{\epsfig{file=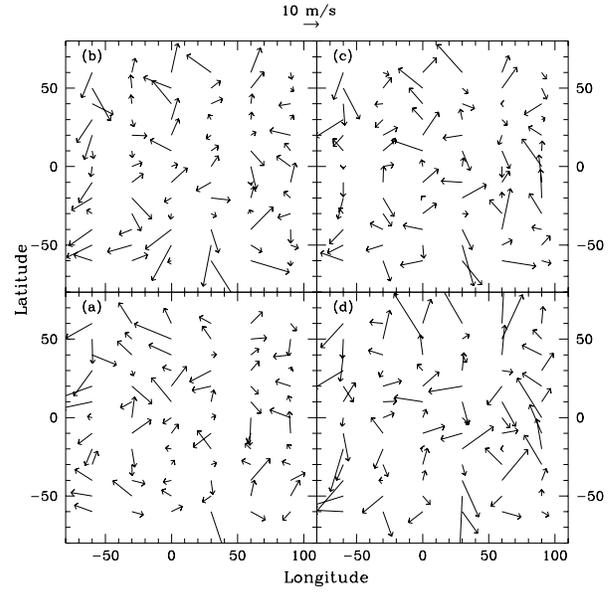,width=8cm}}
  \end{center}
  \caption{\em  Horizontal flow velocities in horizontal planes
at various depths obtained after the rotation and meridional
velocities (as shown in Fig. 6 \& 8) are subtracted from the individual
measurements. The panels (a), (b), (c) and (d) respectively show the
residual flow pattern at radial distance of 0.995, 0.990, 0.985
and $0.980R_\odot$
as inferred by the OLA inversion. The arrow at the top marks the scale.
The errors in these measurements are not shown but they are typically 2--5 m/s
depending on latitude and depth.
}
\end{figure}

In order to extract other components of large scale flow fields and to study
their variation with latitude and longitude we subtract out
the average $u_x$ and $u_y$ as determined from the summed spectra,
from those for individual regions.
The residual velocities are shown in Fig.~10. Some of this velocity could
be due to supergranules. It may be noted that the velocities at
different longitudes in this figure gives the average values
centered at different times. The rms residual velocity at each of these depths
is found to be 12--16 m/s.
There does not appear to be any clear
pattern in this velocity thus suggesting that the giant cells if they
exist have velocities less than about 10 m/s, or their life-time is smaller
than the duration of about 15 days for which the data has been analyzed
in this work, or their longitudinal (latitudinal) size is less than about
$30^\circ$ ($10^\circ$).

\section{CONCLUSIONS}

The average longitudinal velocity agrees reasonably well with
the rotation rate inferred from inversion of global p-modes.
The zonal flow velocity in the outermost region also agrees 
with that estimated from the splitting coefficient for the
global f-modes, though it is not clear if this flow
continues in deeper layers.
The antisymmetric component of the rotation velocity is small
($<5 m/s$) and it is not clear if it is significant.

The dominant signal in the meridional velocity is the meridional
flow which varies with latitude and has a maximum magnitude of about 40 m/s.
The amplitude of meridional component increases from about 25 m/s just
below the surface to 40 m/s in deeper layers. The form of variation with
latitude is not very clear, but the velocity increases with 
latitude until about $40^\circ$ latitude at all depths. There is no
change in sign of meridional velocity with depth up to 21 Mm.

The residual after removing the dominant rotation and meridional
flow component has a magnitude of about 10 m/s and may represent
flows due to the giant
cells or some residual contribution from supergranules.
From the absence of any clear pattern in these flows it appears that
the giant cells if they exist have a velocities less than 10 m/s,
or have lifetimes smaller than 15 days, or their longitudinal size is
less than about $30^\circ$.

\section*{ACKNOWLEDGMENTS}

This work  utilizes data from the Solar Oscillations
Investigation / Michelson Doppler Imager (SOI/MDI) on the Solar
and Heliospheric Observatory (SoHO).  SoHO is a project of
international cooperation between ESA and NASA.
The authors would like to thank the SOI Science Support Center
and the SOI Ring Diagrams Team for assistance in data
processing. The data-processing modules used were
developed by Luiz A. Discher de Sa and Rick Bogart, with
contributions from Irene Gonzalez Hernandez and Peter Giles.

\end{document}